\documentclass[review,authoryear,12pt]{elsarticle}

\usepackage[usenames, dvipsnames]{color}

\newcommand{\Navy}{\color{Black}}
\newcommand{\Orange}{\color{Black}}

\usepackage{psfig}

\usepackage{amssymb}

\journal{PEPI}

\begin{document}

\begin{frontmatter}

\title{A Common Origin for Ridge-and-Trough Terrain on Icy Satellites by Sluggish Lid Convection}

\author[label1,label2,label3]{Amy C. Barr}
\author[label1]{Noah P. Hammond}
 \address[label1]{Department of Geological Sciences, Brown University, Providence, RI 02912, USA}
\address[label2]{RIAMD, Inc., Boulder, CO 80302, USA}
\address[label3]{Planetary Science Institute, Tucson, AZ 85719, USA}

\begin{abstract}
Ridge-and-trough terrain is a common landform on outer Solar System icy satellites.  Examples include Ganymede's grooved terrain, Europa's gray bands, Miranda's coronae, and  {\Orange several terrains on} Enceladus.  {\Orange The} conditions associated with the formation of each of these terrains are similar: heat flows of order tens to a hundred milliwatts per meter squared, and deformation rates of order $10^{-16}$ to $10^{-12}$ s$^{-1}$.   {\Orange Our prior work shows} that the conditions associated with the formation of {\Orange these terrains on Ganymede and the south pole of Enceladus} are consistent with vigorous solid-state ice convection in a shell with a weak surface.  {\Orange We show that sluggish lid convection, an intermediate regime between the isoviscous and stagnant lid regimes,} can create the heat flow and deformation rates appropriate for ridge and trough formation on a number of satellites, {\Orange regardless of the ice shell thickness.}  For convection to deform their surfaces, the ice shells must have yield stresses similar in magnitude to the daily tidal stresses.  {\Orange Tidal} and convective stresses deform the surface, and the spatial pattern of tidal cracking controls the locations of ridge-and-trough terrain.
\end{abstract}


\end{frontmatter}

%
%

\section{Introduction}
Each of our {\Orange Solar System's} outer planets, Jupiter, Saturn, Uranus, and Neptune, harbors a family of regular satellites ranging from $\sim$200 to 2600 km in radius.  Most of these moons have mean densities 1000 kg m$^{-3}< \rho < 2000$ kg m$^{-3}$, in between the values inferred for ice and silicate rock ($\rho_i=$920 kg m$^{-3}$ and $\rho_r =$3000 kg m$^{-3}$), suggesting mixed ice+rock compositions.  Many of the satellites show signs of endogenic resurfacing, some in the form of extensional tectonics, which may occur as organized systems of sub-parallel ridges and troughs, so-called ``ridge-and-trough terrain,'' or as systems of large canyons (e.g., Ithaca Chasma on Tethys), or smaller faults (e.g., ``wispy'' terrain on Dione).  Radiogenic heating supplies a modest heat flow in the interiors of most moons, with surface heat fluxes of order one to tens of mW m$^{-2}$ \citep{SpohnSchubert, Hussmann2006, SchubertEncel2007}.  Thus, much of the endogenic resurfacing is thought to have occurred early in the satellites' histories, when additional sources of heat from accretion, short-lived radioisotopes, and/or differentiation were available to drive resurfacing.  

Organized groups of sub-parallel ridges and troughs, which we refer to as ``ridge-and-trough terrain,'' {\Orange occur} in a number of settings on various satellites.  Examples include Ganymede's grooved terrain (see \citealt{GanymedeJupBook} for discussion), Europa's bands \citep{ProckterBands,Stempel2005}, Miranda's coronae \citep{Pappalardo1997, Hammond-Miranda}, and swaths of ridges and troughs in the northern plains of Enceladus \citep{BlandEncel}.  Each of these terrains are characterized by sub-parallel ridges and troughs with kilometer-scale spacing (see Table 1).  The fault spacing implies a shallow brittle/ductile depth and thus, a high thermal gradient at the time of formation (e.g., \citealt{Nimmo2002}).  Some of the terrains, such as Europa's bands,  subdued grooves on Ganymede, and Elsinore corona on Miranda, are bounded by a sharp groove, and seem to represent sites of emplacement of fresh material from the subsurface (e.g., \citealt{PappalardoSullivan1996, Sullivan98, Head2002}).  Other terrains, such as Arden corona on Miranda and ridge systems on Enceladus appear to be sites of extension without complete lithospheric separation \citep{Pappalardo1997, BlandEncel}.  

Although other satellites show evidence of tectonics, closely spaced sub-parallel ridges and troughs seem to occur primarily on satellites that have experienced tidal flexing, which provides a large heat source and a possible means of lithospheric weakening.  When the orbital periods of satellites within the same system are integer multiples, gravitational interactions between the bodies increase their orbital eccentricities and/or inclinations \citep{MalhotraMiranda,ShowmanMalhotra97, MeyerWisdom2007}, leading to the periodic raising and lowering of a tidal bulge.  If the satellite has a subsurface liquid water ocean, which appears to be common \citep{Zimmer,KivelsonGany, Hussmann2006}, most of the tidal energy is deposited in the outer ice I shell of the satellite, which floats atop the liquid ocean (see Figure \ref{fig:geophysics}a,b) and is free to deform (e.g., \citealt{OS89}).  Through the process of tidal heating, spin energy from the parent planet is converted to mechanical energy via tidal flexing of the satellites' interiors, where it is ultimately dissipated as heat.  Tidally flexed icy satellites can experience heat flows of tens to a hundred mW m$^{-2}$ during their time in resonance (e.g., \citealt{CRPTidal, CRPTidalCorrection, MalhotraMiranda, MeyerWisdom2007, Hammond-Miranda}).  The cyclical deformation of the satellites' outermost ice shells may also weaken the near-surface ice, facilitating deformation \citep{encel_sluggish}.

Solid-state convection has long been suggested to play a role in driving deformation on the icy satellites (e.g., \citealt{Parmentier1982, Shoemaker82, Pappalardo98}).  However, the relationship between convection and resurfacing remains unclear (see, e.g., \citealt{europachap} for discussion). Because the viscosity of water ice depends strongly on temperature \citep{GoldsbyKohlstedt}, convective motions in the satellites' shells are thought to be confined to a relatively thin layer at the base of the shell, beneath a thick Òstagnant lidÓ of cold ice that is too stiff to participate in convection.  {\Orange However, just as stagnant lid convection is not consistent with the relationship between mantle convection and deformation on terrestrial planets like Earth and Venus \citep{MantleConvection}, it does not seem to be consistent with the appearance of Europa, Ganymede, Miranda, or Enceladus, \citep{Parmentier1982, Pappalardo98, Psquared, Stempel2005, HelfLPSC2006, Porco2006, BlandEncel, encel_folds, KattenhornProckter}, nor the measured heat flow of Enceladus \citep{Spencer2006, Howett2011}.}

{\Orange Several recent works have explored the possibility that tidal flexing is weakening and/or heating near-surface ice, permitting convective plumes to reach close to the surface, delivering high heat flows, and driving deformation \citep{ShowmanHan2005, RobertsNimmoMobile, encel_sluggish, Han2012}.  This style of convection, known as the ``sluggish lid'' regime, is an intermediate regime of behavior between isoviscous and stagnant lid convection \citep{Solomatov95, MantleConvection}.  It is characterized by vigorous convection beneath a thin lid of surface material that is dragged along at modest velocities rates by the underlying convective flow \citep{MantleConvection}.}  {\Orange Prior work shows that sluggish lid convection in the shells of Enceladus \citep{encel_sluggish}, Miranda \citep{Hammond-Miranda}, and Ganymede \citep{Hammond-Ganymede} can result in heat fluxes up to $\sim 200$ mW m$^{-2}$ and $\sim~$mm/yr deformation rates, consistent with the conditions of formation for ridge-and-trough terrain on these bodies. } 

Here, we use numerical simulations of ice shell convection to show that similar rheological parameters can give rise to the heat flows and deformation rates inferred for ridge-and-trough formation on four satellites.  We determine scaling relationships for the heat flow and strain rate in extensional regions on the satellites and show that these quantities are {\Orange only weakly dependent on the thickness of the ice shell.}  The yield stresses required for sluggish lid behavior are similar in magnitude to the daily tidal stresses on each satellite.  Tides may weaken the near-surface ice and/or provide lines of pre-existing weakness that can be exploited by convective buoyancy stresses to make swaths of extensional ridges and troughs.  

\section{Observations}
\subsection{Inferred Formation Conditions}
\begin{figure}[h!]
 \centerline{\includegraphics[scale=0.65]{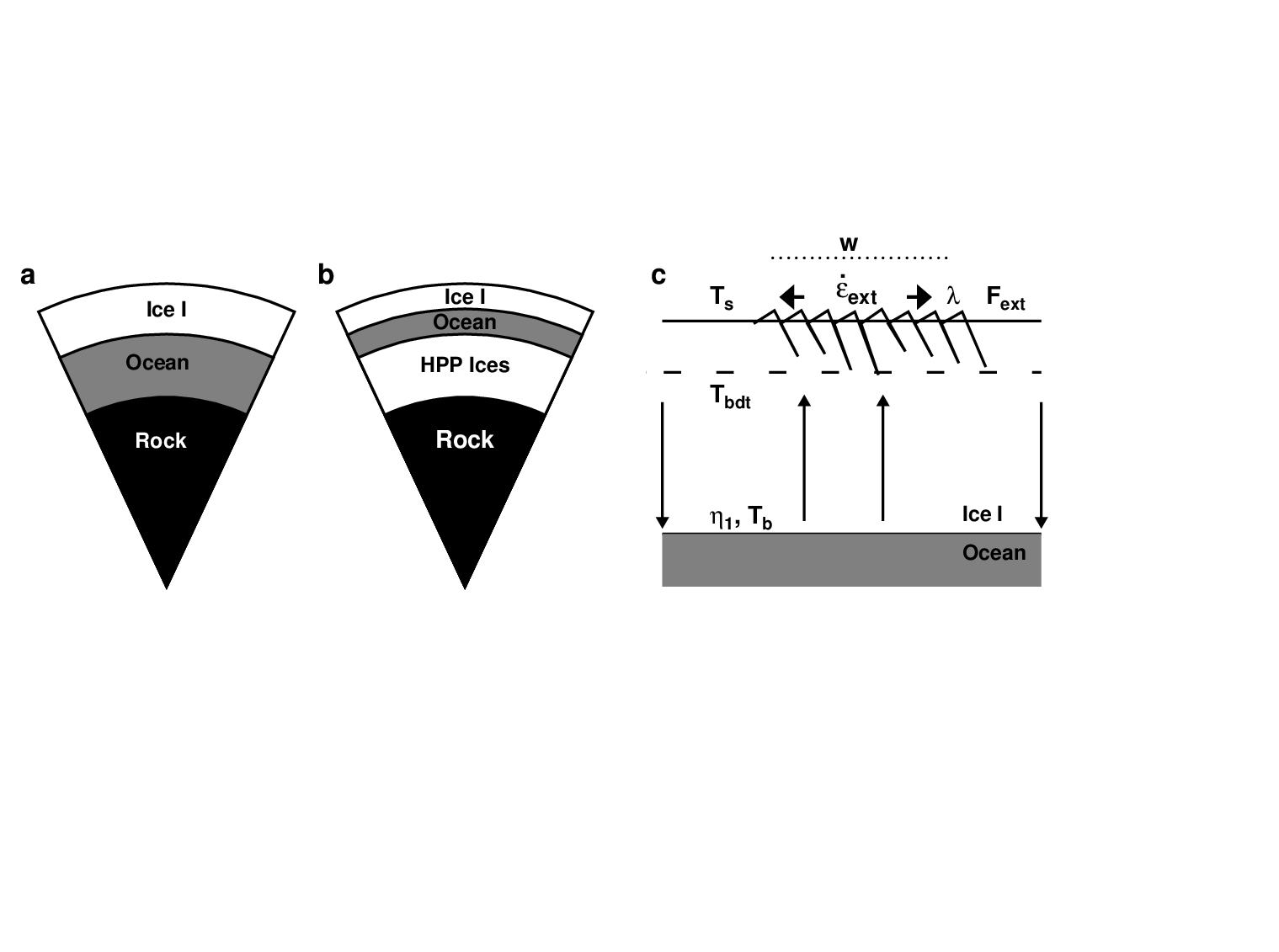}}
     \caption{(a) Schematic (not to scale) of the possible interior structures of Europa, Enceladus, and Miranda.  The satellite has differentiated to form a rocky core, a liquid ocean, and ice I shell.  (b) Schematic representation of Ganymede's interior at the time of grooved terrain formation.  A liquid ocean lies between a layer of ice I and a mantle of high-pressure ice polymorphs.  (c) Schematic (not to scale) illustrating the relationship between deformation 
     and ice I shell convection.  Ridge-and-trough terrain with wavelength $\lambda$ forms in a zone of width $w$, in a near-surface layer due to extensional strain (at strain rate $\dot{\varepsilon}_{ext}$).  The characteristic heat flow ($F_{ext}$) is estimated based on the thickness of the deformed layer, and the difference between the surface temperature $T_s$ and an assumed brittle-ductile-transition temperature ($T_{bdt}$).  Vertical arrows indicate upward/downward convective motion.  Viscosity at the base of the ice shell is $\eta_1$, corresponding to a temperature $T_b$.\label{fig:geophysics}}  
\end{figure}

{\Navy Panels a and b of Figure \ref{fig:geophysics} illustrate the likely interior structures for the bodies in this study.  All four satellites are thought to be fully differentiated, with complete ice/rock separation and formation of a central rock core \citep{Greenberg1991,JupBookInteriors, SchubertEncel2007}.  The subsurface oceans of Europa and Enceladus are thought to be in direct contact with rock.  If an ocean existed on Miranda in the past, it would likely have been in direct contact with the rock core.  Ganymede's ocean is sandwiched between layers of ice I (the low-pressure phase which floats on liquid water), and high-pressure ice polymorphs \citep{JupBookInteriors}.}

There are several methods by which the effective heat flow during deformation can be estimated from characterization of topography on the surface of a planet.  {\Orange One common approach is to derive an estimate of the thickness of the brittle/elastic layer at the surface of the satellite based on measurements of the dominant spacing between ridges and troughs ($\lambda$, see Figure \ref{fig:geophysics}c).  The topography measurements can be performed on digital elevation models, or on images (where brightness is taken as a proxy for topography, e.g., \citealt{Patel1999}).  At the base of the brittle layer, the behavior of the ice is assumed to change from elastic to viscous, at some temperature, $T_{bdt}$.}  Using assumptions about the rheology of ice and the strain rate, lithospheric deformation models can yield estimates of the dominant wavelength of deformation and its relationship to the temperature and depth of the brittle/ductile transition (e.g., \citealt{DombardMcKinnon2001, BlandShowman2007, Bland2010}).  The temperature and depth of the brittle/ductile transition can be used to estimate the thermal gradient and thus, the heat flow (see Figure \ref{fig:geophysics}c).

A second common approach is to estimate the effective elastic thickness of the ice shell by looking for flexural uplift near deformed terrains (e.g., \citealt{Nimmo2002}).  The deformed terrain is assumed to represent a load emplaced on the lithosphere, which drives flexural warping.  The wavelength of the flexural deformation is proportional to the thickness of the elastic portion of the lithosphere.  An assumption of strain rate and ice rheology can yield an estimate of the temperature at which the ice behavior transitions from elastic to viscous.  Similar to the method based on fault spacing described above, the temperature at the base of the elastic layer and the elastic layer thickness permit an estimate of heat flow.  

Table \ref{table:observations} summarizes geological and geophysical constraints on deformation wavelength, heat flux, and strain rate derived from photogeology of several examples of extensional ridge-and-trough terrain.  Here, we mostly use information determined from measurements of fault spacing.  The main sources of error in both the fault spacing and flexural methods are the assumptions about ice rheology and the strain rate.  Although the rheology of the ice in the upper few kilometers can be constrained by laboratory measurements \citep{GoldsbyKohlstedt}, constraints on strain rates are much looser.  Minimum and maximum strain rates are often estimated based on the ages of the terrain (e.g., a feature estimated to be $\tau=1$ Gyr old could have formed with a strain rate as low as $\dot{\varepsilon} \sim 1/\tau \sim 3 \times 10^{-17}$ s$^{-1}$).  Uncertainties in thermal gradient are typically a factor of $\sim 2$, and strain rates may be constrained only to within several orders of magnitude.  These uncertainties are reflected in the values listed in Table \ref{table:observations}.

\subsection{Morphology}
Roughly two thirds of the surface of Jupiter's ice/rock moon Ganymede (with radius $R=$2631 km and mean density $\rho=$1940 kg m$^{-3}$), is covered with bright, relatively young ``grooved terrain,'' swaths of sub-parallel ridges and troughs, often within a sharp bounding groove (see \citealt{GanymedeJupBook} for discussion).  Groove lanes are $\sim10$ to 100 km wide \citep{CollinsChaos} and contain groups of sub-parallel ridges and troughs spaced by $\sim1$-2 km, superimposed on broad pinches and swells $\sim$8 km apart \citep{Patel1999}.  The double wavelength structure of grooved terrain is most readily created by extensional necking at strain rates $\sim10^{-16}$ to $10^{-13}$ s$^{-1}$ in an ice shell with a very high thermal gradient, $\sim5$ to 30 K km$^{-1}$ \citep{BlandShowman2007,Bland2010}.  Flexural studies indicate heat flows at the time of deformation between 80 to 200 mW m$^{-2}$ \citep{Nimmo2002}.  This heat flow is far in excess of that implied by radiogenic heating \citep{DombardMcKinnon2001}, suggesting that the terrain formed during Ganymede's passage through an orbital resonance, which may have also triggered global ice/rock separation and the formation of an ocean \citep{ShowmanStevensonMalhotra}.   Some groove lanes, so-called ``subdued grooves,'' show signs of extensive strike-slip motion and may be sites of emplacement of fresh material from beneath \citep{Head2002}.  

Features called ``bands'' on Jupiter's moon Europa ($R=$1569 km; $\rho=$3040 kg m$^{-3}$) share morphological similarities with the subdued grooves on Ganymede \citep{Head2002} and with terrestrial mid-ocean ridges \citep{ProckterBands}.  Bands are lanes of sub-parallel ridges and troughs roughly 6 to 25 km wide, characterized by a central trough, a hummocky zone, and sets of imbricate fault blocks \citep{ProckterBands} with $\sim$0.5 km spacing \citep{Stempel2005}.  Similar to subdued grooves on Ganymede, bands appear to be sites of complete lithospheric separation and emplacement of fresh material from below \citep{SchenkMcKinnon1989,PappalardoSullivan1996, Sullivan98}.  \citet{Stempel2005} applied a simple model of mid ocean ridge spreading to show that the characteristic fault block spacing in bands can form in an ice shell deforming with strain rates $\sim 10^{-15}$ to $10^{-12}$ s$^{-1}$ with a brittle/ductile transition temperature at $T_{bdt} \sim 150$ to 190 K, at a depth of $2$ to $\sim10$ km.  For a nominal thermal conductivity $k=3.52$ W m$^{-1}$ K$^{-1}$ (see Table \ref{table:satellites}), this corresponds to a heat flow of $15$ to 150 mW m$^{-2}$.  

The surface of Uranus's icy moon Miranda ($R=236$ km; $\rho=1200$ kg m$^{-3}$) is dominated by three zones of intense deformation, dubbed ``coronae.''  
Coronae are $200 - 300$ km in diameter, polygonal to ovoidal in shape and have concentric outer belts of sub-parallel linea \citep{Smith1986}. 
The outer belts surrounding each corona have distinct morphologies.  Arden corona has concentric ridges and troughs with $\approx 5$ km spacing and $2$ km of relief.  Inverness corona has ridges and troughs whose spacing increases with distance from the corona center \citep{Pappalardo1997}.  Elsinore corona has relatively widely spaced ridges and troughs with subdued topography \citep{Schenk1991}. 
The outer belts are interpreted as normal faults and cryovolcanic materials \citep{Greenberg1991, Schenk1991, Pappalardo1997} and each corona is consistent with formation under concentric tensional stresses that radiate from the feature's center \citep{Collins2010}.  
Topography along the flanks of Arden corona suggest the surface may be supported by flexure and that the elastic thickness during corona formation was likely $\approx 2$ km, suggesting a thermal gradient of $8 - 20$ K km$^{-1}$ \citep{Pappalardo1997}.  With a thermal conductivity of 4.2 W m$^{-1}$ K$^{-1}$, this implies a heat flow of 
34 to 84 mW m$^{-2}$.  Such a large thermal gradient could be generated by energy dissipation in Miranda's interior during an orbital resonance with neighboring satellite Umbriel \citep{TittemoreWisdom}. 
Convection driven by tidal heating during resonance can match the distribution of surface deformation and the thermal gradient implied by flexure. 

Saturn's small moon Enceladus has two types of ridge-and-trough terrain.  Its south polar terrain, a $\sim70,000$ km$^2$ quasi-circular region, is a site of intense tectonic deformation and high regional heat flow, with a power output measured by the Cassini CIRS instrument at 3-7 GW \citep{Spencer2006}.  Much of the thermal emission is being emitted from four sub-parallel linear features dubbed ``tiger stripes,'' which are also sites of eruptions of water ice particles with small amounts of silicate and salt (e.g., \citealt{Porco2006,Waite2006, Postberg2009}).  Regions in between the tiger stripes are characterized by sub-parallel folds with a spacing of 1.1 $\pm$ 0.4 km and a funiscular texture, reminiscent of ropy pahoehoe \citep{encel_folds}.  A simple folding model shows that strain rates $10^{-14}$ to $10^{-12}$ s$^{-1}$ can recreate the observed fold spacing \citep{encel_folds}.  Away from the south polar terrain, Enceladus has several large systems of extensional ridge-and-trough terrain with a dominant spacing between faults $\sim3$-4 km \citep{BlandEncel}.  Creating such short-wavelength features requires an effective elastic thickness of only 0.4 to 1.4 km, implying a local heat flow $F\sim110$ to 220 mW m$^{-2}$ at the time of deformation \citep{BlandEncel}, similar to the heat flow currently estimated for the south polar terrain.

\section{Methods}
\subsection{Model}
We simulate convection using the finite element model CITCOM \citep{MoresiSolomatov95}.  The vigor of convection is expressed by the Rayleigh number, 
\begin{equation}
Ra_1=\frac{\rho g \alpha \Delta T D^3}{\kappa \eta_1}, \label{eq:Rayleigh}
\end{equation}
where $\rho$ is the density of ice, $g$ is the acceleration of gravity, $\alpha=1.56 \times 10^{-4} (T_b/250$ K) K$^{-1}$ \citep{KirkStevenson} is the coefficient of thermal expansion, $\Delta T$ is the difference in temperature between the surface ($T_s$) and basal ($T_b$) ice, $D$ is the thickness of the ice shell, $\kappa=1.47 \times 10^{-6} (250\textrm{ K}/T_b)^2$ m s$^{-2}$ is the thermal diffusivity \citep{KirkStevenson}, and $\eta_1$ is the viscosity of ice evaluated at $T=T_b$.  Table \ref{table:satellites} summarizes values of these parameters used for each satellite in the study.  We vary the ice shell thickness between $\sim10$ and 100 km, consistent with estimates for each of the satellites.  We use a temperature dependent Newtonian rheology for ice (cf., \citealt{Solomatov95}), 
\begin{equation}
\eta (T)= \eta_0 \exp (-\gamma T)
\end{equation}
where $\gamma=\theta/\Delta T$ and $\theta=\ln(\Delta \eta)$, where $\Delta \eta = \eta_0/\eta_1$ is the ratio between the viscosity at the surface of the ice shell and its base.  {\Navy We also use the surface Rayleigh number, $Ra_0=Ra_1/\Delta \eta$, to relate the convective heat flow and deformation rates to the thermal and physical properties of the ice shell.}

\subsection{Rheology}
{\Orange Deformation in solid ice is thought to be accommodated by three distinct microphysical mechanisms, diffusion creep at low stresses and in ice with a small grain size \citep{FrostAshby, GoldsbyKohlstedt}, grain-size-sensitive creep, which may occur due to basal slip and grain boundary sliding at intermediate conditions, and dislocation creep at high stresses and in ice with a large grain size \citep{GoldsbyKohlstedt}.  At conditions appropriate for the ice shells of satellites in this study, convective stresses are small enough to be accommodated by diffusion creep \citep{paper2, encelracr}, especially if the ice grain size is kept small by the presence of silicate microparticles intimately mixed in the ice \citep{encelracr,gsevol}.}  We use a nominal value of $\eta_1=3\times10^{14}$ Pa s, {\Orange in the middle of the range commonly assumed in icy satellite convection studies (e.g., \citealt{ShowmanHan2005, NeillNimmo}), and close to the melting point viscosity of ice deforming due to volume diffusion for a grain size $d=0.1$ mm, close to the lower limit of what could be expected in the satellites' ice shells and mantles \citep{encelracr,gsevol}.}{\Navy  If the ice grain size is substantially larger ($d\sim 1$ mm), and deformation is accommodated by non-Newtonian grain size sensitive creep (stress exponent $n\sim 2$) or dislocation creep ($n=4$), the shells of small moons like Enceladus and Miranda are unlikely to convect, due to their low gravity \citep{encelracr}.  Non-Newtonian convection is marginally possible in the thickest ice shells of Europa and Ganymede \citep{paper2}, but only if the grain size of ice is limited to $d \lesssim 1$ mm.  Here, we explore the relationship between convection and resurfacing in a purely Newtonian system; more complex behaviors will be explored in a future study.}  

{\Orange It has long been assumed that ice shell convection could not generate enough stress to overcome the laboratory-derived yield stress of water ice (see e.g., \citealt{SquyresCroft, ShowmanHan2005, europachap} for discussion), and thus, could not drive resurfacing.  Convection in the satellites' ice shells was thought to occur in the stagnant lid regime of convective behavior, requiring extra effects such as compositional buoyancy to drive resurfacing (e.g., \citealt{BarrPappalardo04}).  On terrestrial planets, a promising approach has been to limit the viscosity of the surface material due to the effect of a yield stress (e.g., \citealt{Kohlstedt1995, TackleyScience2000, Bercovici2003, Solomatov2004}).}  

The effect of a finite yield stress in rock and ice is commonly modeled in purely viscous convection models such as CITCOM by setting $\max(\eta)\approx \sigma_Y/\dot{\varepsilon}_{II}$, where $\sigma_Y$ is the yield stress of the convecting material (described by Byerlee's law; \citealt{Beeman}), and $\dot{\varepsilon}_{II}$ is the second invariant of the strain rate tensor (e.g., \citealt{TrompertHansen1998, MoresiSolomatovVenus, ShowmanHan2005, NeillNimmo}).  Here, we use a simpler approach, where we simply limit the viscosity of ice to a constant value, between $10^2$ to $10^5$ times larger than the basal viscosity \citep{encel_sluggish, Hammond-Ganymede, Hammond-Miranda}.  In a Newtonian fluid, sluggish lid behavior occurs for $\Delta \eta \lesssim 10^4$ to $10^5$ \citep{Solomatov95}.  We favor a simpler approach because it limits the number of free parameters in our calculations, allowing us to focus on the relationship between the behavior of the near-surface ice and its rheology.

\subsection{Simulations}
\begin{figure}[h!]
 \centerline{\includegraphics[scale=0.75]{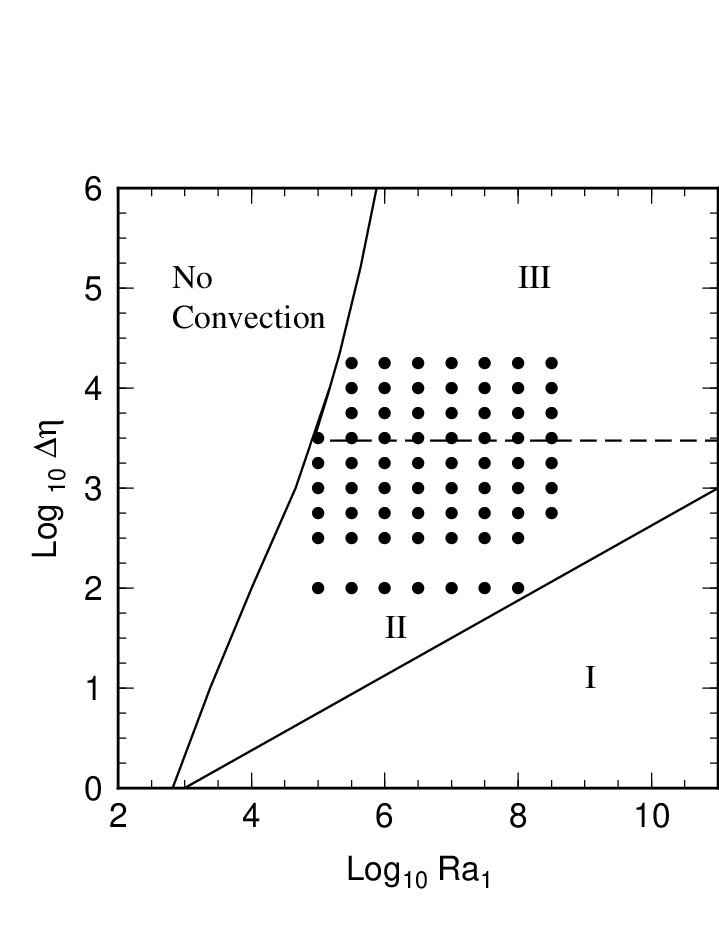}}
     \caption{Locations of our simulations in $Ra$-$\Delta \eta$ space.  Lines show the boundaries between the constant-viscosity convection regime (I),
     transitional (or sluggish lid) regime (II), the stagnant lid regime (III), and no convection.  The boundary between the sluggish and stagnant lid regimes
     (dashed line) is thought to lie near $\Delta \eta \sim 4(n+1) \sim  3000$ for a Newtonian fluid ($n=1$), but its exact location is not precisely defined \citep{Solomatov95}. \label{fig:simlist}}  
\end{figure}

Figure \ref{fig:simlist} illustrates the values of $Ra_1$ and $\Delta \eta$ explored in our study.  Our simulations are performed in an $8 \times 1$ two-dimensional Cartesian domain with $1024 \times 128$ elements, with periodic boundary conditions \citep{Hammond-Ganymede}. The surface and base of the ice shell are held at constant temperatures $T_s$ and $T_b$, implying that the ice shell is heated purely from its base.  Although we imagine that ridge-and-trough terrain is formed in ice shells undergoing active tidal flexing and heating, the details of how the mechanical energy of tidal flexing is converted to heat in the satellites' interiors are not {\Orange well understood.  One possibility is that tidal deformation is converted to energy via Maxwell viscoelastic dissipation, and could be concentrated in the warmest ice \citep{SotinChaos, ShowmanHan2005,MitriShowmanTidal2008}.  Another possibility is that cyclical tidal deformation along shallow ice faults could dissipate energy, thus concentrating tidal heating in the coldest ice \citep{NimmoGaidos}.  Exploring other distributions of tidal heat is a promising direction of future study.}

\subsection{Outputs}
Simulations are run until the dimensionless heat flow, or Nusselt number ($Nu$) has reached a statistical steady state, so that 
\begin{equation}
Nu_{rms}= \frac{\int_0^t (Nu(t)) dt}{\int_0^t dt}
\end{equation}
has converged to the $10^{-5}$ level \citep{SM2000}.  We measure the strain rate, $\dot{\varepsilon}_x=dv_{x,sf}/dx$ , where $v_{x,sf}$ is the $x-$velocity at the surface.  The surface experiences extension where $\dot{\varepsilon}_x>0$.  In these regions, we record the average heat flow, $Nu_{ext}$, and the strain rate, $\dot{\varepsilon}_{ext}$.  Successful simulations are those in which both of these quantities match the values in Table 1.  

\section{Results}
\subsection{Heat Flow}
The heat flow across the convecting ice shell is related to the Nusselt number,
\begin{equation}
F=\frac{k \Delta T}{D} Nu,
\end{equation}
where $k=651/T$ W m$^{-1}$ K$^{-1}$ is the temperature-dependent thermal conductivity of ice {\Orange \citep{PW}}, which we evaluate at $T=T_s+\frac{1}{2}\Delta T$ (see Table \ref{table:satellites}).  {\Orange In general, $Nu \varpropto Ra^b$, where the value of $b$ is constrained by numerical simulations, where $Ra$ can be $Ra_1$ or $Ra_0$, depending on what is more convenient for the problem at hand.} 

\subsubsection{Prior Work}
{\Navy Many prior works have sought to constrain the value of $b$ in Cartesian and 3-dimensional spherical domains, for both constant- and variable-viscosity convection (e.g., \citealt{BercoviciJFM1989, Solomatov95, SotinLabrosse1999, DeschampsSotinGJI2000, SM2000, Wolstencroft2009, Deschamps2010}).  For isoviscous convection in a spherical shell, \citet{Wolstencroft2009} find $Nu \varpropto Ra^{0.294}$ for basally heated shells, and $Nu \varpropto Ra^{0.337}$ for internal heating, where differences in the power law index $b$ are attributed to differences in convective planform.  For similar conditions, \citet{Deschamps2010} find $Nu \varpropto Ra^{0.273+0.05f}$, where $f$ is a non-dimensional ratio between the core radius and the planetary radius.  In two- and three-dimensional Cartesian domains, with both internal and basal heating, $b=1/3$ \citep{SotinLabrosse1999, SM2000}. Early studies, including the first groundbreaking efforts to estimate $b$ in spherical shells or with strongly temperature-dependent viscosity, found $b \sim 0.25$ to $0.28$ \citep{BercoviciJFM1989, Solomatov95, DeschampsSotinGJI2000}.  However, these works had a tendency to underestimate $b$ owing to computational difficulties (e.g., limited resolution and domain geometry).  For example, \citet{BercoviciJFM1989} characterize $b$ for a single harmonic mode.  The first studies to estimate $b$ in 1$\times1$ Cartesian boxes (e.g., \citealt{Solomatov95,DeschampsSotinGJI2000}) used relatively low numerical resolution and tended to underestimate $Nu$ in vigorously convecting systems, leading to artificially low values of $b$.}

{\Orange Because sluggish lid convection can be viewed as a transitional regime between isoviscous and stagnant lid convection, \citet{OlsonCorcos} propose $b=1/3$.  \citet{encel_sluggish} used numerical simulations in Cartesian geometry, in a $1 \times 1$ box with free-slip boundary conditions, to characterize the $Ra-Nu$ relationship for the sluggish lid regime.  \citet{encel_sluggish} proposed a scaling of form (cf. \citealt{MoresiSolomatovVenus}),}
\begin{equation}
Nu = a Ra_0^{b} \exp(\theta/c), \label{eq:scaling}
\end{equation}
{\Orange where  $Ra_0$ is the Rayleigh number evaluated using $\eta=\eta(T_s)$, which can be calculated from the basal Rayleigh number, $Ra_0 = Ra_1/\Delta \eta$.  Assuming $b \equiv 1/3$, \citet{encel_sluggish} found $a=0.32$ and $c=19$.}

\subsubsection{Nusselt Number} 
\begin{figure}[h!]
 \centerline{\includegraphics[scale=0.75]{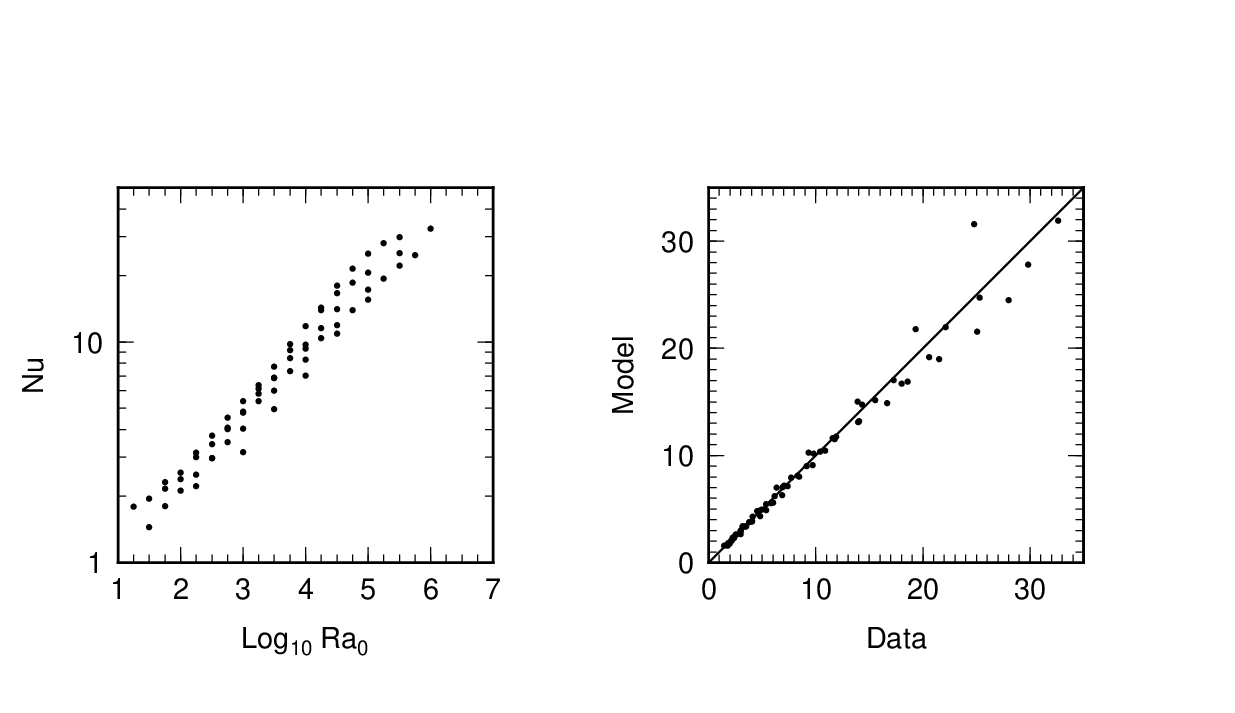}}
     \caption{(left) Dimensionless heat flow, $Nu$, from our simulations as a function of surface Rayleigh number, $Ra_0$. (right) Comparison between
     the values of $Nu$ from our simulations (``data'') with the scaling relationship (equation \ref{eq:scaling}; ``model''). \label{fig:Nu}}  
\end{figure}

{\Navy Here, we improve upon the results of \citet{encel_sluggish} by simulating convection in a wider box, for a wider range of parameters, and at higher resolution.  To determine the relationship among $Ra_0$, $Nu$, and $\theta$, we allow $a, b,$ and $c$ to be free parameters, and estimate their values based on a best fit to data from our simulations.  The left panel of Figure \ref{fig:Nu} shows how the values of $Nu$ obtained in our simulations vary as a function of $Ra_0$.  We find that $b$ remains constant over several orders of magnitude in Rayleigh number.  We note that we slightly underestimate $Nu$ at high $Ra$, which is likely due to our resolution.  The $Ra-Nu$ relationship close to the critical Rayleigh number has a different dependence, likely $b \approx 1/2$, which would be expected for extremely low-amplitude convection \citep{SolomatovBarr2007}.  A multivariate least squares fit on all of the $Nu$ values in our data set gives $a=0.23 \pm 0.03$, $b=0.32 \pm 0.007$, and $c=9.8 \pm 0.12$.  The right panel of Figure \ref{fig:Nu} shows how equation (\ref{eq:scaling}) compares to the data.}
%

\subsubsection{Heat Flow in Extensional Zones}
\begin{figure}[h!]
 \centerline{\includegraphics[scale=0.85]{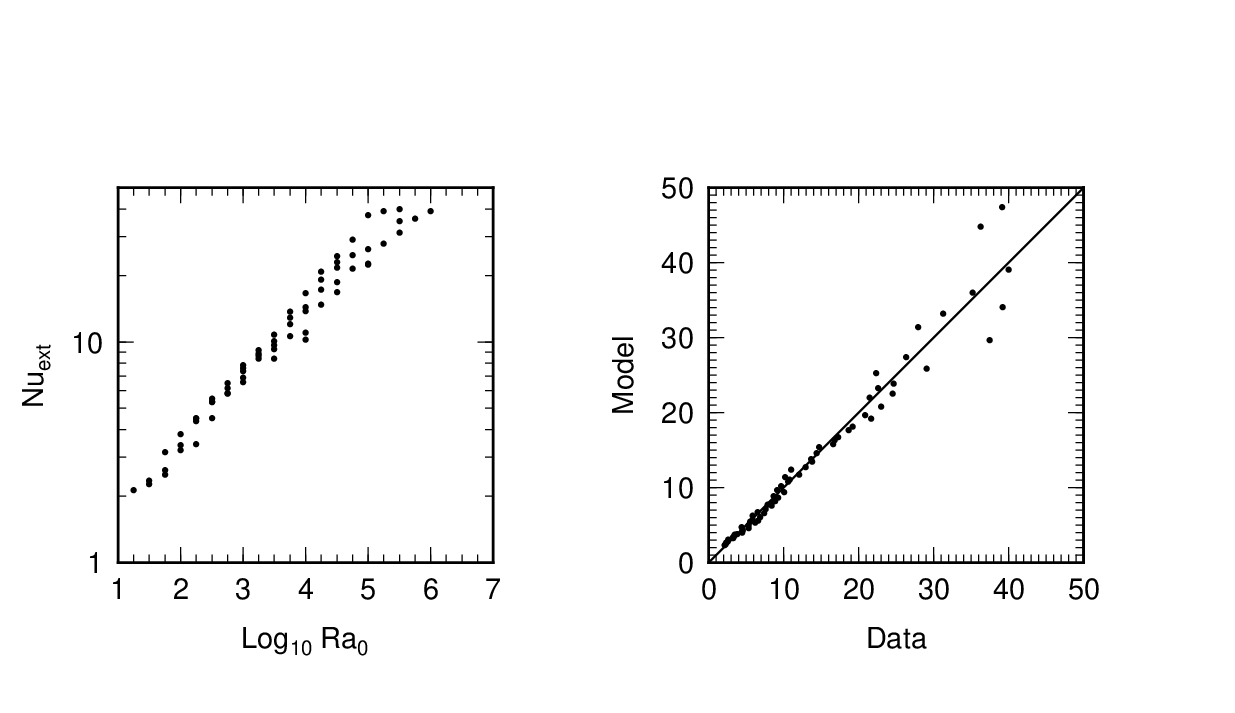}}
     \caption{(left) Dimensionless heat flow in extensional zones, $Nu_{ext}$, from our simulations as a function of surface Rayleigh number, $Ra_0$. (right) Comparison between
     the values of $Nu_{ext}$ from our simulations with the scaling relationship (equation \ref{eq:ext_Nu}). \label{fig:ext_Nu}}  
\end{figure}

{\Orange To determine whether sluggish lid convection can create extensional ridge-and-trough terrain, we also need to determine the relationship between physical properties of the ice shell and the heat flow in extensional regions, $Nu_{ext}$.  Figure \ref{fig:ext_Nu} summarizes the values of $Nu_{ext}$ obtained in our study, and their relationship with $Ra_0$.  One might expect that the relationship between this quantity and the Rayleigh number might be similar to equation (\ref{eq:scaling}).  Indeed, a fit to our numerical results, allowing $a,b$, and $c$ to be free parameters, shows that $Nu_{ext}$ can be described by,
\begin{equation}
Nu_{ext} = (0.48 \pm 0.09) Ra^{0.31 \pm 0.009} \exp\bigg(\frac{\theta}{14.21 \pm 2.7}\bigg), \label{eq:ext_Nu}
\end{equation}
which is extremely similar to the scaling for $Nu$, with roughly a factor of $\sim 2$ difference between the two.  }
%

\subsubsection{Implications for Resurfacing}
\begin{figure}[h!]
 \centerline{\includegraphics[scale=0.85]{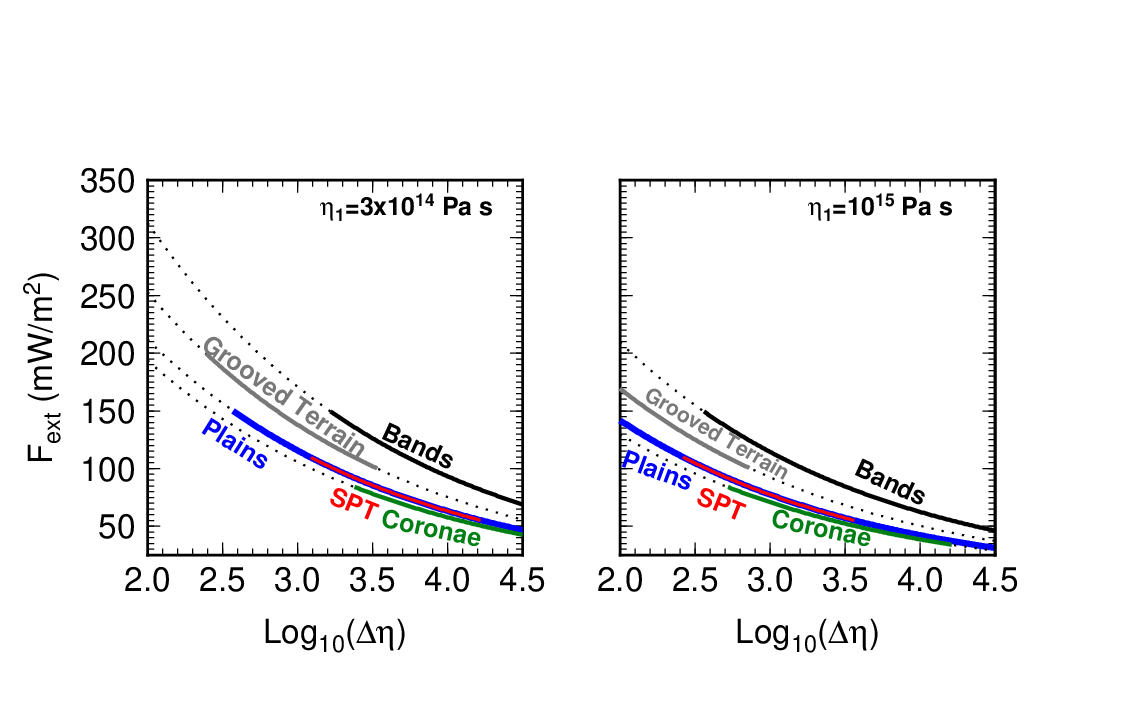}}
     \caption{(left) Heat flow in regions of extension $F_{ext}$, for $\eta_1=3 \times 10^{14}$ Pa s, as a function of $\Delta \eta$ for Europa (black), Ganymede (gray), Enceladus (blue/red), and Miranda (green).  Dotted lines indicate values of $\Delta \eta$ where heat flows are outside the range inferred from geological characterization.  Solid lines indicate values of $\Delta \eta$ where heat flows are consistent with the formation of each of the class of surface features we consider: Europa's bands, Ganymede's grooved terrain, Miranda's coronae; and the south polar terrain and fold systems in the northern plains of Enceladus. (right) Same as (left), but for $\eta_1=10^{15}$ Pa s.\label{fig:Fext.eps}}  
\end{figure}

Figure \ref{fig:Fext.eps} illustrates the range of $\Delta \eta$ values for which $F_{ext}$ matches geological constraints for the formation of Europa's bands, Miranda's coronae, Enceladus' South Polar Terrain, the fold belts in the northern plains of Enceladus, and Ganymede's grooved terrain.  {\Orange For $\eta_1=3 \times 10^{14}$ Pa s,} we find that ice shells with $10^{2.5} < \Delta \eta < 10^{4.25}$ can create heat flows high enough to create each of these features.  Ice shells with $10^{3.25} < \Delta \eta < 10^{3.5}$ are consistent with the creation of all five sets of features we consider.  This range of $\Delta \eta$ is also consistent with the heat flow and morphology of convective upwellings required to create the global tetrahedral distribution of Miranda's coronae \citep{Hammond-Miranda}.  {\Orange The range of $\eta_1$ commonly used in icy satellite convection studies is $\eta_1 \sim 10^{13}$ Pa s to $10^{15}$ Pa s.  If $\eta_1 =10^{13}$ Pa s, deformation rates and heat flows are higher than those observed on the icy satellites.  However, $\eta_1 =10^{13}$ corresponds to an ice grain size $d \sim$ 1 micron, not physically plausible in a natural system \citep{gsevol}.  Larger values of $\eta_1$ (see right panel of Figure \ref{fig:Fext.eps}) provide a broader range of $\Delta \eta$ in which the conditions of ridge-and-trough terrain formation can occur.}

{\Navy Because $b\approx 1/3$, we do not expect the heat flow in extensional zones to depend strongly on the thickness of the ice shell, $D$.  However, simulations of isoviscous convection in a spherical shell show that $b$ may be slightly larger or smaller than $1/3$ (e.g., \citealt{Wolstencroft2009}), giving rise to a heat flow that would be very weakly dependent on the thickness of the ice shell (i.e., $D^{0.01}$).  Given the very weak dependence on $D$, and uncertainties in other properties of the system (chiefly, $\eta_1$), it is not possible to obtain a robust estimate of ice shell thickness based on the heat flow in extensional zones.  However, on a small satellite such as Miranda, where curvature affects the pattern of upwellings and downwellings \citep{Deschamps2010, Hammond-Miranda}, it is possible to get loose constraints on $D$ by comparing the \emph{global} pattern of ridge-and-trough terrain zones with the convection pattern in a three-dimensional spherical shell \citep{Hammond-Miranda}.}



\subsection{Deformation Rates}
\subsubsection{Prior Work}
{\Orange Similar to the Nusselt number, r.m.s. convective velocities depend on the Rayleigh number and $\theta$, and are thought to follow a scaling relationship of form, (cf. \citealt{SM2000}) $u = a_u \theta^{-\alpha_u} Ra_i^{\beta_u},$ where $\alpha_u=\beta_u$, and the values of the fitting coefficients are constrained by results from numerical simulations.  Boundary layer theory applied to basally heated convection suggests $\beta_u=2/3$ for Newtonian convection (see e.g., \citealt{TurcotteSchubert, Solomatov95} for discussion).  However, an alternative scaling derived from equating viscous dissipation in the rheological sublayer with work per unit time done by buoyancy forces gives $\beta_u=0.5$ \citep{SM2000}.  Isoviscous convection simulations in a three-dimensional spherical shell indicate $\beta_u=0.538$ for basal heating and $\beta_u=0.5$ for internal heating \citep{Wolstencroft2009}.}  

To determine whether convective strain rates are likely to form extensional ridge-and-trough terrain on icy satellites, we are primarily interested in a different quantity: the maximum strain rate in extensional zones at the surface, which is related to the maximum convective velocity at the surface, $\max(|v_{x,sf}|)$.  {\Navy Our prior work finds that $\max(|v_{x,sf}|) \approx a_v Ra_0^{b_v}(\kappa/D)$, where $a_v=0.08$ and $b_v=0.8$ \citep{encel_sluggish}, with no statistically significant dependence on $\theta$. }

\subsubsection{Maximum Deformation Rate}
\begin{figure}[h!]
 \centerline{\includegraphics[scale=0.85]{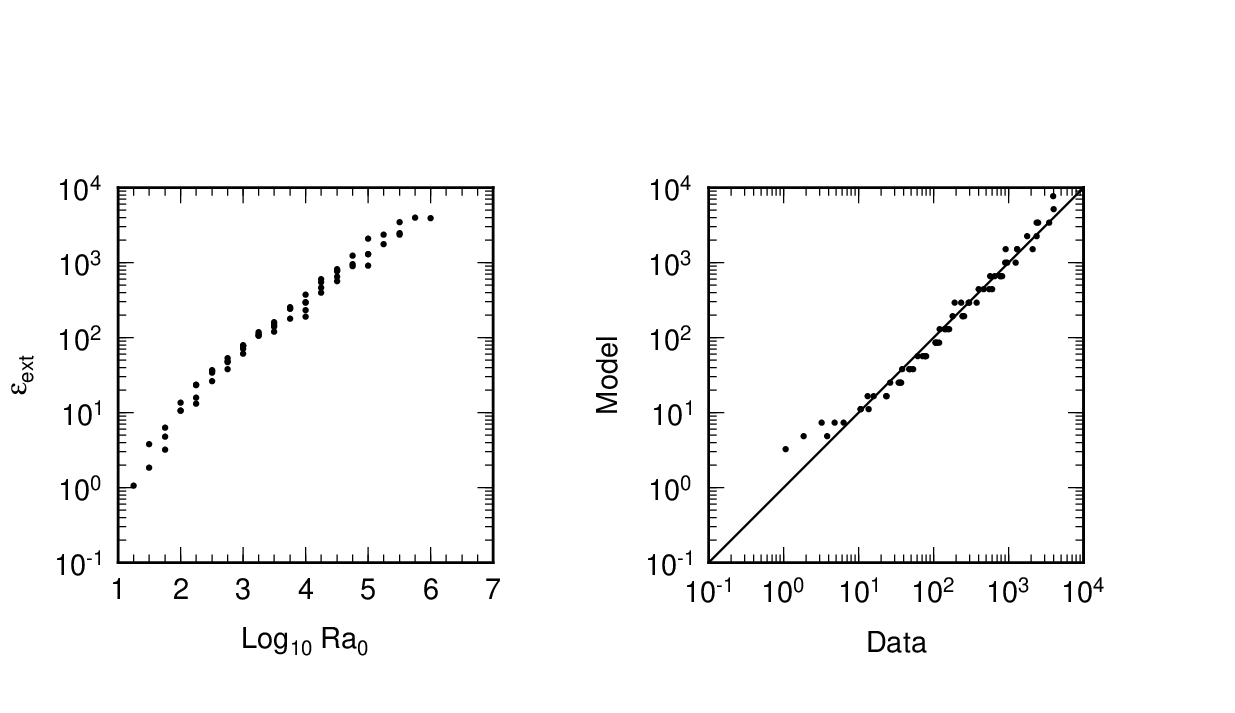}}
     \caption{(left) Dimensionless values of strain rate in extensional zones, $\dot{\varepsilon}_{ext}$ from our simulations as a function of surface Rayleigh number, $Ra_0$. (right) Comparison between
     the values of $\dot{\varepsilon}_{ext}$ from our simulations (``data'') with the scaling relationship (equation \ref{eq:vx_scaling}; ``model''). \label{fig:vx}}  
\end{figure}

{\Orange Here, we expand on our prior study to constrain the maximum strain rate occurring on the surface of the convecting layer, $\dot{\varepsilon}_{ext}$, which we expect to follow a relationship similar to that for $\max(|v_{x,sf}|)$,
\begin{equation}
\dot{\varepsilon}_{ext} = a_e Ra_0^{b_e} \frac{\kappa}{D}. \label{eq:vx_scaling}
\end{equation}
Fits to our data give $a_e=0.42 \pm 0.12$ and $b_e=0.71 \pm 0.03$.  A three-parameter fit including dependence on $\theta$ yields similar values for $a_e$ and $b_e$, but with an additional term $\theta^{0.05 \pm 0.05}$.  Because the error bars encompass a zero value for the power on $\theta$, we cannot state that $\dot{\varepsilon}_{ext}$ depends on $\theta$ with a reasonable degree of statistical confidence.  Figure \ref{fig:vx} illustrates the values of $\dot{\varepsilon}_{ext}$ obtained in our study, and a comparison between these values and the two-parameter fit given in equation (\ref{eq:vx_scaling}).  For the same reasons we have errors in $Nu$ at low and high $Ra_0$, we find some evidence of systematic error in $\dot{\varepsilon}_{ext}$.  }

\subsubsection{Implications for Deformation}
\begin{figure}[h!]
 \centerline{\includegraphics[scale=0.55]{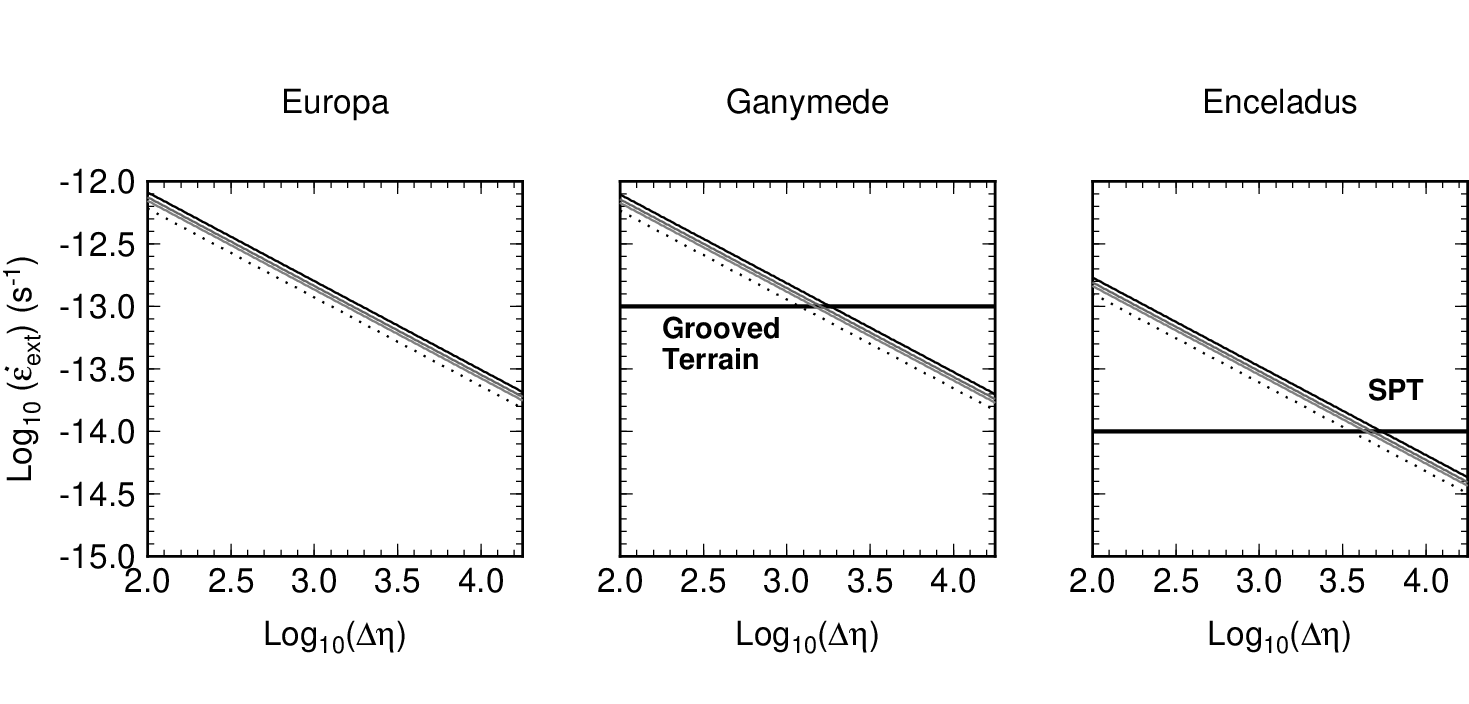}}
     \caption{Strain rate in extensional regions ($\dot{\varepsilon}_{ext}$) as a function of viscosity contrast ($\Delta \eta$) for Europa (left), Ganymede (middle), and Enceladus (right).  Strain rates depend weakly on ice shell thickness ($D$) and are reported for a range of values: $D=10$ km (dotted), $D=30$ km (light gray), $D=50$ km (dark gray), and $D=100$ km (black).   Across all $\Delta \eta$ values in our study, strain rates in extensional regions match those for Europa's bands.  Strain rates associated with grooved terrain formation are achieved for $\Delta \eta \gtrsim 10^{3.25}$.  On Enceladus, strain rates in the south polar terrain imply $\Delta \eta < 10^{3.75}$ at that location, but the north polar extensional features may form for any $\Delta \eta$ in our study.  \label{fig:Vext.eps}}  
\end{figure}

Figure \ref{fig:Vext.eps} illustrates how the strain rates in extensional zones predicted by equation (\ref{eq:vx_scaling}) vary as a function of $\Delta \eta$ and the thickness of the ice shells on Europa, Ganymede, and Enceladus.  Strain rates associated with the formation of Europa's bands are readily obtained for a wide range of $\Delta \eta$.  The low strain rates associated with grooved terrain formation imply $\Delta \eta \gtrsim 10^{3.25}$.  Strain rates predicted for the Enceladus SPT can be created only in very weak shells with $\Delta \eta < 10^{3.75}$, but strain rates associated with the formation of ridges and troughs in the northern plains are created for a wide range of $\Delta \eta$.  

{\Navy Although the relationship between $Ra$ and $Nu$ has been relatively well constrained, the relationship between convective velocities and $Ra$ is less certain (see e.g., \citealt{SM2000} for discussion).  With the scaling proposed in equation (\ref{eq:vx_scaling}), $\dot{\varepsilon}_{ext} \varpropto D^{3b_e - 2}$, which gives, for our value of $b_e$, $\dot{\varepsilon}_{ext} \varpropto D^{0.13}$.  If $b_e=0.5$ \citep{SM2000}, $\dot{\varepsilon}_{ext} \varpropto D^{-0.5}$, which could allow a rough estimate of the ice shell thickness based on extensional strain rates.  More simulations, in more complex geometries, with higher numerical resolution, and for a broader range of $\Delta \eta$ values, could provide a better estimate of $b_e$ and might allow a constraint on ice shell thickness based on estimated deformation rates in extensional zones.}

\section{Discussion}
Extensional ridge-and-trough terrain is a common landform on tidally flexed icy satellites.  Examples include the grooved terrain on Ganymede, bands on Europa, Miranda's coronae, and ridges and troughs in the northern plains of Enceladus.  On each of these satellites, the spacing between ridges and troughs is of order a kilometer, up to ten kilometers.  Each of these terrains is inferred to form in an ice shell with a high thermal gradient, and thus, heat flow.  Strain rates between about $10^{-16}$ and $10^{-12}$ s$^{-1}$ (see Table \ref{table:observations}) are consistent with the spacing between ridges and troughs observed.  

Our numerical simulations of convection in an ice shell with a weak surface show that the heat flow and strain rate associated with ridge-and-trough terrain can be created by sluggish lid convection.  This conclusion holds on each satellite regardless of the thickness of the ice shell because the heat flow and strain rate are {\Orange only very weakly dependent on $D$.  However, if the satellite's ice shell occupies a significant fraction of its radius (possibly the case for Miranda), curvature will affect the global convection pattern of upwellings and downwellings, and the global distribution of resurfaced areas could shed light on the shell thickness \citep{Hammond-Miranda}.}

{\Orange For reasonable melting point viscosities for ice, we find that a single set of rheological parameters can give rise to conditions appropriate for the formation of \emph{all} of the terrains.}  By limiting the value of $\Delta \eta$ to less than $10^4$ to $10^5$, we are mimicking the effect of weak near-surface ice.  A crude estimate of the ice strength in our model, $\sigma_y \sim \eta_0 \dot{\varepsilon}_{ext}$, implies that the near-surface ice must have a yield strength $\sim 1$ to 300 kPa for sluggish lid behavior.  This yield stress is many orders of magnitude lower than that inferred for the yield stress of ice based on terrestrial field studies \citep{Kehle1964} and laboratory experiments \citep{Beeman}.  However, the formation of the cycloidal cracks on Europa \citep{HurfordCycloids2007}, the timing and duration of plume eruptions on Enceladus \citep{HurfordNature2007}, and the putative eruptions on Europa \citep{Roth2014}, suggest that the modest daily tidal stresses exerted on these bodies are capable of cracking near-surface ice.  These tidal stresses are of similar magnitude to $\sigma_y$ \citep{HurfordCycloids2007, HurfordNature2007, SatStress}.  

This suggests that the stresses from tides and the stresses from solid-state convection are both required to create ridge-and-trough terrain.  The cyclical tidal flexing of the ice shells has been suggested to produce a significant amount of heat in the shells via tidal dissipation (e.g., \citealt{OS89}).  However, we suggest that the cyclical flexing is also modifying the structure of the near-surface ice, at the macro- and possibly micro-scale.  

At macro-scales, pre-existing lines of weakness driven by tidal forces may provide weak locations in the ice shell where thermal buoyancy stresses from underlying convection can pull the surface apart.  This may explain why only tidally flexed icy satellites have global occurrences of extensional ridge-and-trough terrain, whereas satellites of similar size and composition that have not experienced tidal flexing have little endogenic activity.  Classic examples of this apparent paradox include Ganymede (tidally flexed) and Callisto (no tidal flexing and no endogenic resurfacing); and Enceladus (tidally flexed) and Mimas (no tidal flexing and no endogenic resurfacing).  If convection can only drive deformation along lines of weakness, this might explain, for example, the global degree-2 pattern of grooved terrain on Ganymede \citep{Patterson2010}.  Convection may also create ridge-and-trough terrain by pulling apart pre-existing tidal cracks in the satellites' ice shells.  On Europa, reconstruction of pre-existing features across bands suggests that these features represent locations of the emplacement of new material \citep{PappalardoSullivan1996, Sullivan98}.  When the regions of fresh material are removed, the remaining landform resembles a simple ubiquitous europan crack in the ice shell (see, e.g., Figure 2 of \citealt{PappalardoSullivan1996}).  

At micro-scales, the cyclical working of the ice shell at low temperatures (where kinetic processes such as grain growth and annealing are slow compared to the timescale of tidal flexing), may introduce cracks and defects in the ice that severely weakens its structure, decreasing its yield stress.  Similar processes, along with the presence of pore fluids, may be responsible for decreasing the yield stress of the crust on the Earth, permitting plate tectonics (see, e.g., \citealt{Kohlstedt1995} for discussion).  Another possibility is that tidal flexing may introduce heat in the near-surface ice, providing thermal softening \citep{RobertsNimmoMobile} which allows for intense deformation driven by convection.  In either case, laboratory experiments are needed to clarify the response of ice to cyclical flexing at frequencies and temperature conditions appropriate for the ice mantles of the outer planet satellites.  

Here, we have focused on the role of convection in driving, principally, extensional deformation on the icy satellites.  Definitive morphological evidence for compression on icy satellites is rare (e.g., \citealt{Psquared}).  However, recent work by \citet{Bland2012} suggests that low-amplitude compressional folds can form at strain rates and thermal gradients similar to conditions arising in the compressional zones in our simulations \citep{Hammond-Ganymede}.  Thus, it may be possible that the seemingly ubiquitous extension on icy satellites may be accommodated by folds that are difficult to detect in existing images.  Further images from spacecraft, e.g., the forthcoming {\Orange NASA Europa and ESA Ganymede missions}, could shed light on the relationship between extension and compression on icy bodies.

\section*{Acknowledgements}
This work was supported by NASA OPR NNX12AL22G and NESSF NNX13AN99H.  We thank an anonymous reviewer and Editor Mark Jellinek for valuable comments.


%
%
\clearpage
\begin{table}[h!]
\begin{tabular}{l l | cccc}
\hline
Satellite & Feature & $\lambda$ (km) & $w$ (km) & $F$ (mW m$^{-2}$) & $\dot{\varepsilon}$ (s$^{-1}$)  \\
\hline
Enceladus & South Polar Terrain$^{a}$ & 1.1 $\pm$0.4 & -- & 55--110& 10$^{-14}$ -- 10$^{-12}$ \\
Europa & Bands$^{b,c}$ & 0.5 & 6--25 & 15--150& 10$^{-15}$ -- $10^{-12}$ \\
Enceladus & Northern Plains$^d$ & 3--4 & 30--50 & 30--150 & 10$^{-15}$ -- $10^{-12}$ \\
Ganymede & Grooved Terrain$^{e,f,g,h}$ & 2; 8 & 10--100 & 100--200 & 10$^{-16}$ -- $10^{-13}$ \\
Miranda & Coronae$^{i}$ & 5--10 & -- & 34 -- 84 & -- \\
\hline
\end{tabular}
\caption{Wavelength of deformation ($\lambda$), width of deformed zones ($w$), inferred heat flow ($F$), and strain rates ($\dot{\varepsilon}$) estimates for ridge-and-trough terrains on various icy satellites.  $^a$\citet{encel_folds}, $^b$\citet{ProckterBands}, $^c$\citet{Stempel2005}, $^d$\citet{BlandEncel}, $^e$\citet{Nimmo2002}, $^f$\citet{Patel1999}, $^g$\citet{DombardMcKinnon2001}, $^h$\citet{BlandShowman2007}, $^i$\citet{Pappalardo1997}. \label{table:observations}}
\end{table}

\clearpage
\begin{table}[h!]
\begin{tabular}{l l | c c c c}
\hline
Parameter & Symbol & Europa & Ganymede & Enceladus  & Miranda \\
\hline
Surface Temperature 	& $T_s$ (K) 		& 110 & 130 &	80   & 60 \\
Basal Temperature 	& $T_b$ (K) 		& 260 & 260 &	273  & 250 \\
Ice Shell Density 		& $\rho$ (kg m$^{-3}$) & 920 & 920  &	920 & 920  \\
Gravity 			& $g$ (m s$^{-2}$)  	& 1.3 & 1.42 &	0.11 &  0.079 \\
Ice Thermal Expansion & $\alpha$ (K$^{-1}$) 	&$1.62 \times 10^{-4}$ & $1.62 \times 10^{-4}$  &	$1.70 \times 10^{-4}$  & $1.56 \times 10^{-4}$ \\
Thermal Diffusivity 	& $\kappa$ (m s$^{-2}$) 	& $1.36 \times 10^{-6}$ & $1.36 \times 10^{-6}$ &	$1.23 \times 10^{-6}$ & $1.47 \times 10^{-6}$  \\
Basal Viscosity 		& $\eta$ (Pa s)			& $3 \times 10^{14}$ & $3 \times 10^{14}$ & $3 \times 10^{14}$  &  $3 \times 10^{14}$ \\
Thermal Conductivity   & $k$ (W m$^{-1}$ K$^{-1}$)			& 3.52 & 3.34 & 3.69 & 4.2  \\
\hline
\end{tabular}
\caption{Thermal and physical properties of satellite ice shells.  \label{table:satellites}}
\end{table}

\clearpage
\clearpage

\clearpage

\clearpage

\clearpage

\clearpage

\clearpage

\end{document}